# Structural and superconducting properties in LaFeAs$_{1-x}$Sb$_x$O$_{1-y}$F$_y$

WANG Cao, MA Zhifeng, JIANG Shuai, LI Yuke, XU Zhu'an, CAO Guanghan[†]

Department of Physics, Zhejiang University, Hangzhou 310027, China

**We report the antimony (Sb) doping effect in a prototype system of iron-based superconductors LaFeAsO$_{1-y}$F$_y$ (y=0, 0.1, 0.15). X-ray powder diffraction indicates that the lattice parameters increase with Sb content within the doping limit. Rietveld structural refinements show that, with the partial substitution of Sb for As, while the thickness of the Fe$_2$As$_2$ layers increases significantly, that of the La$_2$O$_2$ layers shrinks simultaneously. So a negative chemical pressure is indeed "applied" to the superconducting-active Fe$_2$As$_2$ layers, in contrast to the effect of positive chemical pressure by the phosphorus doping. Electrical resistance and magnetic susceptibility measurements indicate that, while the Sb doping hardly influences the SDW anomaly in LaFeAsO, it recovers SDW order for the optimally-doped sample of y=0.1. In the meantime, the superconducting transition temperature can be raised up to 30 K in LaFeAs$_{1-x}$Sb$_x$O$_{1-y}$F$_y$ with x=0.1 and y=0.15. The Sb doping effects are discussed in term of both $J_1$–$J_2$ model and Fermi Surface (FS) nesting scenario.**

superconductivity, crystal structure, doping effect, negative chemical pressure

## 1 Introduction

The recent discovery of LaFeAsO$_{1-x}$F$_x$[1] triggered a frantic gold rush for a new series of high $T_c$ superconductors, evoking old memories of the cuprate superconductors which were the rage more than 20 years ago[2]. Similar to the cuprates, chemical doping is still the commonest strategy to induce superconductivity in the iron-arsenic compounds. So far, take $Ln$FeAsO ($Ln$ stands for lanthanide) for instance, superconductivity can be induced by both hole doping[3] and electron doping[4-9], and the transition temperature has been pushed up to 56 K[6].

Crystallizing in ZrCuSiAs-type[10], LaFeAsO is one of the prototype parent compound of iron-based superconductor, which consists of conducting [Fe$_2$As$_2$]$^{2-}$ layers and insulating [La$_2$O$_2$]$^{2+}$ layers. Neutron diffraction experiment showed that, this parent compound undergoes an antiferromagnetic spin-density-wave (SDW) ordering below ~137 K[11]. When doping electrons, the SDW order is suppressed and then superconductivity emerges. Typically, for LaFeAsO$_{1-y}$F$_y$ system, the superconducting window in the electronic phase diagram is $y \gtrsim 0.05$, and the optimal doping level is around $y=0.1$[1, 12]. When La is replaced by other smaller lanthanide elements, the superconducting window moves to higher doping levels[4, 13-16]. In addition to the chemical doping, superconductivity can also be induced by applying pressures[17].

We previously reported that an isovalent doping by phosphorus (P), which produces chemical pressure in the arsenide system, induced superconductivity in EuFe$_2$As$_{2-x}$P$_x$[18] and La-

[†] Corresponding author (email: ghcao@zju.edu.cn)
This work is supported by the NSF of China (Contract No. 90922002), the National Basic Research Program of China (Contract No. 2007CB925001) and the PCSIRT of the Ministry of Education of China (IRT0754).

$FeAs_{1-x}P_xO$[19] systems. Take $LaFeAs_{1-x}P_xO$ for instance, on the one hand, the partial substitution of As by P shrinks the distance from the P/As atoms to the iron planes, which leads to the evolution of one of the Fermi Surface (FS) pocket from two-dimensional $d_{xy}$ to three-dimensional $d_{3z^2-r^2}$ character[20]. On the other hand, the substitution of P for As results in variations of effective exchange couplings especially for the next-nearest ($J_2$) case which is a superexchange via P/As in nature[21-23]. So no matter which scenario ("FS nesting" or "$J_1$–$J_2$ coupling") is adopted, the modification in the crystal structure plays an important role in both suppressing SDW order and inducing superconductivity.

In this letter we report the Sb doping effect in $LaFeAsO_{1-y}F_y$ system ($y$=0, 0.1, 0.15). X-ray powder diffraction and Rietveld structural refinements show that, Sb doping thickens the $Fe_2As_2$ and thins the $La_2O_2$ layers, exerting *negative chemical pressure* to the superconducting-active $Fe_2As_2$ layers, in contrast to the effect of *positive* chemical pressure by the P doping. Electrical resistance and magnetic susceptibility measurements indicate that, while the Sb doping alone can hardly influence the SDW anomaly in $LaFeAs_{1-x}Sb_xO$, it recovers SDW order for the $LaFeAs_{1-x}Sb_xO_{0.9}F_{0.1}$ samples. On the other hand, the superconducting transition n temperature can be enhanced up to 30 K in $LaFeAs_{0.9}Sb_{0.1}O_{0.85}F_{0.15}$. It is emphasized that, this *negative* pressure can hardly be exerted on the sample through any physical strategy. So this work might be a unique complementarity to reveal how lattice expansion influences the SDW ordering and superconductivity in the so-called "1111" system.

## 2  Experiments

$LaFeAs_{1-x}Sb_xO_{1-y}F_y$ polycrystalline samples were synthesized by solid state reaction in vacuum using powders of LaAs, $La_2O_3$, $LaF_3$, FeAs, $Fe_2As$, FeSb and $Fe_2Sb$. Similar to our previous report[19], LaAs, FeAs, $Fe_2As$, FeSb and $Fe_2Sb$ were presynthesized respectively. $La_2O_3$ was dried by firing in air at 1173 K for 24 hours prior to using. All the starting materials are with high purity (≥99.9%). Powders of these intermediate materials were weighed according to the stoichiometric ratios of $LaFeAs_{1-x}Sb_xO_{1-y}F_y$, thoroughly mixed in an agate mortar, and pressed into pellets under a pressure of 6000 kg/cm$^2$, operating in a glove box filled with high-purity argon. The pellets were sealed in evacuated quartz tubes, then heated uniformly at 1373 K for 50 hours, and finally furnace-cooled to room temperature.

Powder X-ray diffraction (XRD) was performed at room temperature using a D/Max-rA diffractometer with Cu-$K\alpha$ radiation and a graphite monochromator. The detailed structural parameters were obtained by Rietveld refinements, using the step-scan XRD data with $10° \leq 2\theta \leq 120°$. The electrical resistivity was measured using a standard four-terminal method. The measurements of dc magnetic properties were performed on a Quantum Design Magnetic Property Measurement System (MPMS-5). Zero-field cooling (ZFC) protocols were employed under the field of 10 Oe to determine the magnetic shielding percentage.

## 3  Result and discussions

The crystallographic parameters are obtained by the Rietveld refinement based on the ZrCuSiAs-type structure. An example of the refinements is seen in Fig. 1. The reliability factor $R_{wp}$ is 11.0% and the goodness of fit is 1.58, indicating fairly good refinement for the crystallographic parameters. Table 1 lists the structural data of $LaFeAs_{1-x}Sb_xO_{1-y}F_y$ samples. Compared with the undoped LaFeAsO, the *a*-axis increases by 0.84% while *c*-axis increases by 1.52% for LaFeAs$_{0.67}$Sb$_{0.33}$O. As for the samples of $y$=0.1, Sb doping expands the *a*-axis and *c*-axis by 0.75% and 1.33% respectively ($x$=0.2 and $y$=0.1). The obvious change in lattice parameters indicate that the Sb atoms successfully occupy the As site. Besides, the change in structure details attracts our attention. Firstly, for the samples with the same fluorine content ($y$=0, 0.1), it is clear that the As/Sb atoms are getting away from the Fe plane as Sb is doped, resulting in the thickening of the Fe$_2$As$_2$ layers. Surprisingly, the La atoms move toward the oxygen plane, causing the flattening of the La$_2$O$_2$ layers. Thus it is quite obvious that, the Sb doping exerts *negative chemical pressure* on the Fe$_2$As$_2$ layers, contrary to the P doping case[19]. Secondly, with the thickening of the Fe$_2$As$_2$ layers, the Fe-As-Fe angle decrease monotonously, which may account for the enhanced superconducting $T_c$ in LaFeAs$_{0.9}$Sb$_{0.1}$O$_{0.85}$F$_{0.15}$ sample, according to the empirical relationship between the structure and superconducting $T_c$ in iron-arsenic compound[24, 25].

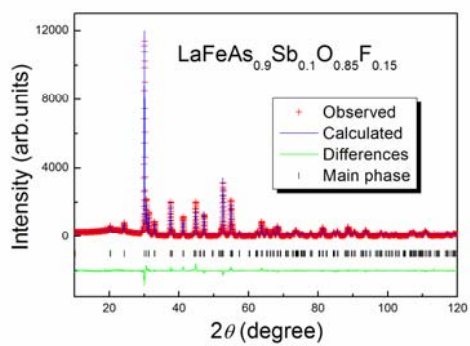

**Fig. 1**. An example of Rietveld refinement profile for LaFeAs$_{0.9}$Sb$_{0.1}$O$_{0.85}$F$_{0.15}$.

The temperature dependence of resistivity for LaFeAs$_{1-x}$Sb$_x$O ($x$=0, 0.33) is shown in Fig. 2. The data are normalized to 300 K as the measured resistivity on a polycrystalline sample is often sensitive to grain boundary and surface effects. The resistivity of the parent compound LaFeAsO shows an anomaly at about 150 K, where structure phase transition occurs. When 33% of As atoms are replaced by Sb, the anomaly is still robust, and no superconductivity is observed above 3 K. Higher doping level leads to a substantial increase in impurity phase. Thus it is obvious that, Sb doping alone can neither suppress SDW order nor induce superconductivity in LaFeAs$_{1-x}$Sb$_x$O system. The inset of Fig. 2 shows the derivative of resistivity (d$\rho$/d$T$) as a function of temperature. Based on the method proposed by Klauss *et al*[26], the critical temperatures associate with the structure phase transition ($T_S$) and SDW ordering ($T_N$) can be recognized as $T_S$=157 K (144 K), $T_N$=138 K (133 K) for $x$=0 ($x$=0.33), respectively. The decrease in the temperature difference between $T_S$ and $T_N$ (from 19 K to 11 K) is consistent with the thinning of the La$_2$O$_2$ layers, which may strengthen the interlayer magnetic exchange coupling significantly[27, 28].

**Table 1.** Crystallographic data of LaFeAs$_{1-x}$Sb$_x$O$_{1-y}$F$_y$ at room temperature. The space group is P4/nmm. The atomic coordinates are as follows: La (0.25, 0.25, $z$); Fe (0.75, 0.25, 0.5); As/Sb (0.25, 0.25, $z$); O/F (0.75, 0.25, 0).

|  | $x$=0.33,$y$=0 | $x$=0,$y$=0 | $x$=0,$y$=0.1 | $x$=0.1,$y$=0.1 | $x$=0.15,$y$=0.1 | $x$=0.2,$y$=0.1 | *$x$=0.1,$y$=0.15* |
|---|---|---|---|---|---|---|---|
| $a$ (Å) | 4.0698(4) | 4.0357(3) | 4.0273(1) | 4.0419(2) | 4.0399(2) | 4.0575(2) | 4.0402(2) |
| $c$ (Å) | 8.8708(9) | 8.7378(6) | 8.7068(3) | 8.7618(4) | 8.7537(4) | 8.8225(5) | 8.7539(4) |
| $V$ (Å$^3$) | 146.93(3) | 142.31(2) | 141.214(7) | 143.13(2) | 142.87(1) | 145.25(1) | 142.89(3) |
| $z$ of La | 0.1380(2) | 0.1411(2) | 0.1473(1) | 0.1433(1) | 0.1434(1) | 0.1399(1) | 0.1437(1) |
| $z$ of As | 0.6550(3) | 0.6513(3) | 0.6511(2) | 0.6532(2) | 0.6541(2) | 0.6562(2) | 0.6533(2) |
| La$_2$O$_2$ thickness (Å) | 2.447(3) | 2.466(2) | 2.565(1) | 2.512(2) | 2.510(2) | 2.469(2) | 2.515(2) |
| Fe$_2$As$_2$ thickness (Å) | 2.750(3) | 2.644(2) | 2.631(2) | 2.684(2) | 2.699(3) | 2.757(2) | 2.684(1) |
| Fe-Fe spacing (Å) | 2.8778(3) | 2.8536(3) | 2.8477(1) | 2.8580(1) | 2.8567(1) | 2.8691(2) | 2.8569(1) |
| Fe-As-Fe angle (°) | 111.9(2) | 113.5(1) | 113.7(1) | 112.8(1) | 112.5(1) | 111.6(1) | 112.8(1) |

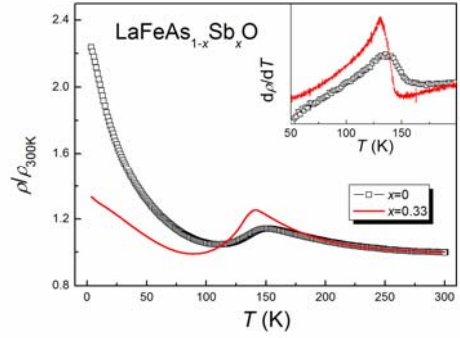

**Fig. 2.** Temperature dependence of resistivity for LaFeAs$_{1-x}$Sb$_x$O samples. The data are normalized for comparison. The inset shows an expanded differential slop d$\rho$/d$T$ to show the SDW anomaly clearly.

To study whether Sb doping will influence the superconducting transition temperature, we take the fluorine doped LaFeAsO$_{0.9}$F$_{0.1}$ superconductor as the starting compound. As is shown in Fig. 3, the $\rho$-$T$ curve of LaFeAsO$_{0.9}$F$_{0.1}$ sample shows a sharp superconducting transition at about 25 K, which is consistent with the previous report[1]. When 10% of the As atoms are replaced by Sb atoms, the superconducting $T_c^{onset}$ is enhanced to 27 K. Interestingly, a low-temperature upturn emerges below 60 K, which is reminiscent of the similar upturn of the underdoped LaFeAsO$_{1-y}$F$_y$[12]. A further increase of Sb content ($x$=0.15, 0.2) leads to a disappearance of superconductivity and a recovery of SDW anomaly, manifesting the competition between the SDW order and superconductivity. For the $x$=0.1/$y$=0.15 sample, the resistivity upturn below 60 K disappears, and the superconducting $T_c^{onset}$ is enhanced to 30 K. The result is consistent with a previous report[29], which shows an enhanced $T_c^{onset}$=30.1 K in LaFeAs$_{1-x}$Sb$_x$O$_{0.8}$F$_{0.2}$ with $x$=0.05.

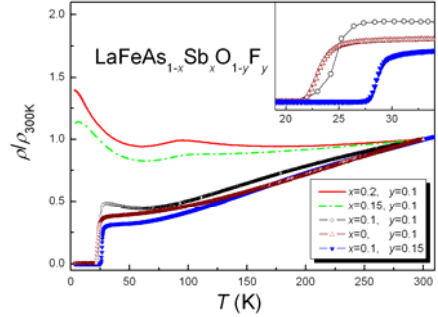

**Fig. 3**. Temperature dependence of resistivity for LaFeAs$_{1-x}$Sb$_x$O$_{1-y}$F$_y$ samples. The inset shows an expanded plot. The data are normalized for comparison.

The superconducting transition in LaFeAs$_{1-x}$Sb$_x$O$_{1-y}$F$_y$ samples is confirmed by dc magnetic susceptibility, shown in Fig. 4. For the $x=0/y=0.1$ and $x=0.1/y=0.1$ samples, strong diamagnetic signal can be seen below 23 K and 24.5 K respectively, which is consistent with the resistance measurements. When the Sb doping level is raised to 20%, no diamagnetic signal is detected above 2 K. As for the $x=0.1/y=0.15$ sample, the superconducting $T_c$ is enhanced to 27 K, and the volume fraction (assuming a theoretical density) of magnetic shielding achieves over 100 % (the demagnetization factor is not taken into account), indicating bulk superconductivity.

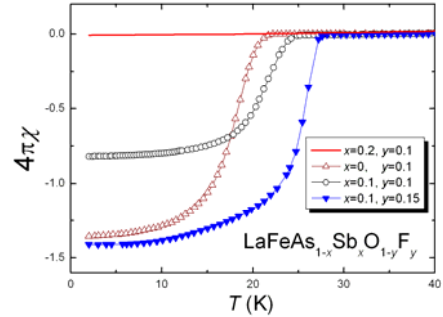

**Fig. 4**. Temperature dependence of dc magnetic susceptibility for LaFeAs$_{1-x}$Sb$_x$O$_{1-y}$F$_y$ samples. The zero-field-cooling protocols are performed at 10 Oe.

Now let's discuss the Sb doping effects in LaFeAs$_{1-x}$Sb$_x$O$_{1-y}$F$_y$ system. Fixing the fluorine doping level (say, $y=0$, 0.1), one can find in Table 1 that, the Sb doping leads to a *negative chemical pressure* effects with squeezed La$_2$O$_2$ layers and stretched Fe$_2$As$_2$ layers along the $c$-axis respectively, which is just opposite to the chemical pressure effect in P doped LaFeAs$_{1-x}$P$_x$O[19]. This may be explained in term of difference in electronegativity for P, As and Sb, since the Sb atoms do not pull La/Fe atoms as much as P atoms do. Furthermore, the increase of As/Sb height relative to the Fe plane results in a monotonous decrease in the diagonal Fe-As-Fe angle. Previous theoretical work indicated that, the Fe-Fe nearest-neighbor ($J_1$) and the next-nearest-neighbor effective exchange couplings ($J_2$) are mediated through the electron Fe–As–Fe hopping, controlled by the Fe–As–Fe angle[21-23]. So the decrease in diagonal Fe-As-Fe angle may enhance the antiferromagnetic coupling and then suppress superconductivity due to the competitive relationship between the two ground states. As shown in Fig. 3, indeed we observed suppression of $T_c$ and recovery of SDW order in LaFeAs$_{1-x}$Sb$_x$O$_{0.9}$F$_{0.1}$. On the other hand, the decrease in the diagonal Fe-As-Fe angle tends to selectively increase $J_2$, which may account for the enhanced $T_c$ for the sample of $x$=0.1/$y$=0.15, according to a theoretical work[30].

A related phenomenon is that the Sb doping modifies the electronic phase diagram. Taking LaFeAs$_{0.9}$Sb$_{0.1}$O as the *starting compound*, the $T_c^{onset}$ for the sample of $y$=0.15 is 3 K higher than that of $y$=0.1. This may indicate that, Sb doping will move the superconducting window in the electronic phase diagram to higher fluorine doping level, which is consistent with the difference in electronic phase diagram for *Re*FeAsO$_{1-x}$F$_x$ (*Re*= Ce, Pr, Nd, Sm, Gd)[4, 13-16] in term of the decrease in diagonal Fe-As-Fe angle.

In the FS nesting scenario, the Sb doping effect can also be explained based on modification in crystal structure. Previous ARPES results indicate that, two hole pockets center at the Γ point and another two electron pockets locate at the M point for LaFeAsO parent compound[31]. Theoretically, when the As atoms approach the Fe plane, all the FS remain nearly unchanged except that one of the two-dimensional (2D) hole pocket with d$_{xy}$ character evolves into a three-dimensional pocket with d$_{3z^2-r^2}$ character[20]. When the As/Sb atoms slightly shift away from Fe plane, it is thus reasonable to speculate that the 2D d$_{xy}$ hole Fermi pocket will remain. So if the SDW ordering is ascribed to the FS nesting between the 2D inner (outer) hole pockets centering at the Γ point and the 2D inner (outer) electron pockets centering at the M point respectively[32, 33], no wonder Sb doping alone can hardly suppress the SDW order in LaFeAs$_{1-x}$Sb$_x$O. For the fluorine doped LaFeAs$_{1-x}$Sb$_x$O$_{0.9}$F$_{0.1}$, the expanded distance from As/Sb to Fe plane may stabilize the d$_{xy}$ hole pocket. Thus the SDW ordering shows up again in the co-doped samples.

## 4  Conclusion

In summary, we have synthesized a series of LaFeAs$_{1-x}$Sb$_x$O$_{1-y}$F$_y$ samples. For the fluorine-free samples, Sb doping alone can neither suppress SDW order nor induce superconductivity in LaFeAs$_{1-x}$Sb$_x$O system. Yet, the d$\rho$/d$T$ curve indicates a decrease in temperature gap between $T_S$ and $T_N$, which is ascribed to the strengthening of the interlayer magnetic exchange coupling. For the co-doped samples, we observe recovery of SDW order in LaFeAs$_{1-x}$Sb$_x$O$_{0.9}$F$_{0.1}$ and enhanced $T_c^{onset}$ up to 30 K in LaFeAs$_{0.9}$Sb$_{0.1}$O$_{0.85}$F$_{0.15}$. Based on XRD measurements and Rietveld refinements, the Sb doping effects in LaFeAs$_{1-x}$Sb$_x$O$_{1-y}$F$_y$ system can be understood in term of both $J_1$–$J_2$ coupling and FS nesting mechanism. Our results suggest that, not only the carrier doping levels, but also the structural details of FeAs layers control the appearance of SDW ordering and/or superconductivity.